\documentclass[%
 reprint,
superscriptaddress,
amsmath,amssymb,
]{revtex4-2}

\usepackage{graphicx}
\usepackage{dcolumn}
\usepackage{bm}
\usepackage{natbib}

\begin{document}


\title{Condensation induced internal convection of two neighboring droplets}

\author{Tapan Kumar Pradhan}
\email{tapan.k.pradhan@gmail.com}
 \affiliation{Department of Mechanical Engineering, \\ Indian Institute of Technology Kharagpur, Kharagpur 721302, India.} 
 
\author{Pradipta Kumar Panigrahi}%
 
\affiliation{Department of Mechanical Engineering, \\ Indian Institute of Technology Kanpur, Kanpur 208016, India.}


\begin{abstract}
The present work investigates the internal flow structure of two condensing droplets of aqueous solution during interaction. The velocity of fluid inside the droplets was investigated by confocal micro-PIV technique. Condensation on the droplets was carried out inside a closed chamber by creating a difference in vapor pressure  between the droplet interface and a reservoir fluid surrounding the droplet in the room temperature without any cooling. Condensation on the droplet leads to spatial variation of solute concentration inside the droplet causing buoyancy driven Rayleigh convection. Fluid flow pattern inside a single condensing droplet is symmetric in nature where as the fluid flow pattern for two interacting droplets is asymmetric in nature. The neighboring droplet influences the internal convection through vapor mediated interaction between the two condensing droplets. Here, the condensing droplet senses the neighboring droplet from a distance without any physical contact. The asymmetric nature of the flow pattern is attributed to the modification in the distribution of condensation flux of the droplet in the presence of the neighboring droplet. The flow pattern observed for interacting droplets during condensation is opposite to that of evaporating droplets. The effect of the neighboring droplet on the internal convection of the condensing droplet reduces with increase in separation distance between the droplets.
\end{abstract}

\maketitle


\section{Introduction}
Presence of a neighboring droplet near to another droplet is very common in natural phenomena and engineering applications like drop wise condensation, microfluidics devices, dew formation etc. Hence, it is very necessary to study the effect of a neighboring droplet on the internal convection of a droplet. The droplet may influence the behavior of the neighboring droplet without having any physical contact between the droplets. Droplets may interact with each other through vapor mediated interaction \cite{Carles1989,Pradhan2015,Pradhan2016,Wray2020} or through substrate mediated interaction \cite{Karpitschka2016,Pandey2017,Sadafi2019}. In vapor mediated droplet interaction, both the droplets sense each other through the vapor cloud of the two droplets. In case of substrate mediated droplet interaction, the droplets sense each other by elastic deformation of the soft substrate on which the droplets are placed. Vapor mediated droplet interaction has drawn significant attention in the recent years. 

\par
Droplets may attract or repel each other through substrate mediated or vapor mediated interaction. The substrate mediated droplet attraction and repulsion has been studied by \citet{Karpitschka2016} where the droplets are placed on a soft surface. They mentioned that the interaction between the droplets occurs by inverted Cheerios effect where elastic deformation of the soft surface leads to attraction or repulsion between the two droplets. The attraction or repulsion between the two droplets depends on the substrate thickness. Similar attraction and repulsion between droplets has also been observed in case of vapor mediated droplet interaction. \citet{Carles1989} studied the movement of a PDMS droplet and a trans-decaline droplet placed on a glass slide. The droplets show movement as a result of variation in surface tension at the two ends of the droplet caused by vapor transport between the droplets. Such attraction and repulsion between the droplets are also being observed for water-propylene glycol mixture placed on a glass surface \cite{Cira2015}. The interaction between the droplets by the vapor medium leads to random motion of the droplets on the surface. Similar attraction or repulsion between droplet pair is observed due to differential evaporation from the droplet surface \cite{Man2017}. Aqueous droplets at the interface of an immiscible liquid pool can show attraction due to the vapor interaction between the droplet through the immiscible liquid medium \cite{Liu2018}. A neighboring droplet can also affect the coffee ring deposition pattern of a drying droplet \cite{Pradhan2015} where the proximity region of the droplets shows weak deposition. The modification of the deposition pattern by a neighboring droplet occurs by the alternation of flow pattern due to vapor mediated droplets interaction. 

\par
Study of internal flow pattern of droplets is very informative to understand the hydrodynamics of droplets. Internal hydrodynamics of interacting droplets has been explored for different cases in last few years. \citet{Pradhan2015} experimentally studied the internal capillary flow of two interacting water droplets placed on a hydrophilic glass slide. The flow pattern shows weak strength at the proximity region as less evaporation is observed at this region. Later, they \cite{Pradhan2016} studied the influence of a neighboring droplet on the internal fluid flow pattern of an evaporating aqueous solution droplet placed on a hydrophobic surface. The neighboring droplet creates asymmetric evaporative flux distribution which is the cause of flow alternation by the neighboring droplet. Droplet interaction is also observed inside microchannel of droplet based microfluidics devices. When two droplets of aqueous solution at different concentrations are placed inside a microchannel, flow is induced inside both the droplets due to vapor transport between the droplets \cite{Pradhan2018a}.  A preliminary study on the effect of a neighboring water droplet on the fluid flow inside an ethylene glycol droplet is reported by \citet{Pradhan2017}. They reported that the vapor emitted from the neighboring water droplet penetrates at the interface of the ethylene glycol droplet which induces fluid flow inside the ethylene glycol droplet. A recent study \cite{Singh2020} on the flow pattern induced inside a sessile droplet containing titania powder particles is affected by the the presence of neighboring similar droplets. The alignment of the titania particles inside the droplets also depends on the number of neighboring droplets surrounding the droplet.  Also internal convection plays a significant role during droplets coalescence \cite{Aversana1996,Karpitschka2010,Sykes2020,Rostami2020}. 

\par
All the droplet interaction phenomena studied by previous researchers are focused on evaporating droplets only. In our previous work \cite{Pradhan2018b}, we reported the internal fluid flow of a single condensing droplet. Hydrodynamics of condensing droplets during interaction has not been explored. In this work, we have explored the vapor mediated interaction of two condensing droplets of aqueous solution where the droplets sense other droplet through modification of condensing flux around the droplet. We experimentally study the phenomena and simulation to understand the underlying physics associated with it.

\section{Experimental Section}
In this work, fluid velocity inside two neighboring condensing droplets was measured using confocal micro-PIV. Internal convection of condensing droplets was studied for both single and two droplets configurations. Condensation on the droplets was carried out inside a condensing chamber in the ambient temperature without any cooling of the droplets. Details of the condensation method and velocity measurement inside the droplets are presented in subsequent subsections.

\subsection{Method of condensation}
\begin{figure}[htb!]
\begin{center}
\includegraphics[width=0.47\textwidth]{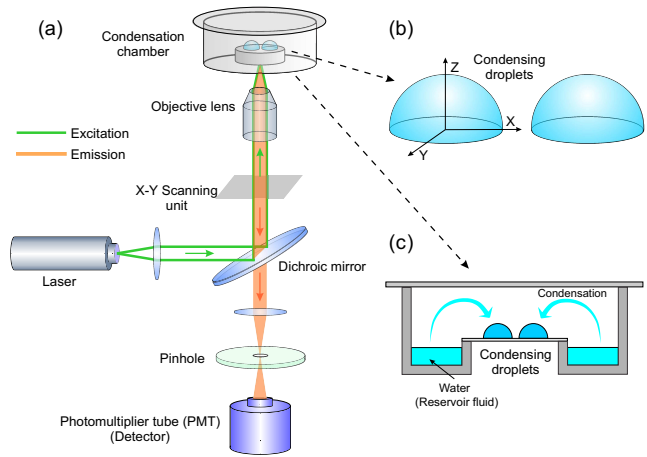}
\caption{\label{fig:exp} Experimental setup for condensation and study of internal hydrodynamics of droplets. (a) Confocal micro-PIV setup for velocity measurement, (b) two droplets of aqueous solution separated by a separation distance and (c) condensation chamber.}
\end{center}
\end{figure}

The condensing chamber was made of polymethyl methacrylate (PMMA). It is cylindrical in shape having internal diameter of 36 mm. The droplets were placed on a circular glass cover slip whose diameter is 22 mm. The glass cover slip was made hydrophobic by surface modification using OTS \cite{Wong2013}. The glass cover slip was mounted on a cylindrical stage having diameter 22 mm fabricated eccentrically to the condensing chamber as shown in the Figs. \ref{fig:exp}(a) and (c). Reservoir fluid (water) of volume 1.5 ml was poured into the annular groove surrounding the central cylindrical stage. After placing the droplets on the cover slip and the reservoir fluid inside the groove, the chamber was sealed from the top by a circular cover. Silicone vacuum grease (Dow Corning Vacuum Grease) was used to glue the cover slip and the top cover to the chamber. This completely isolated the inside chamber environment from the surroundings. Then the condensing chamber was placed on the stage of a confocal microscope for the experiment. The whole setup was allowed to rest for 5 minutes after sealing the condensing chamber to reach a steady state flow condition inside the droplets. 

\par
Two droplets of aqueous NaCl solution of 1 M concentration were used in the experiment. The volume of each droplet was kept at 1 $\mu$L. Each droplet shows a contact angle of \(92^0 \, \pm \, 0^0 \) and forms a circular contact line having diameter of 1.56 mm. The two droplets are separated at a distance of 290 $\mu$m from each other. The droplets were surrounded by a reservoir fluid (water) as shown in Fig. \ref{fig:exp}(c). According to Raoult's law, solute concentration affects the vapor pressure at the interface of a solution. The vapor pressure at the liquid-air interface decreases with increase in the solute concentration.  The vapor pressure difference at the interface of the droplet and the reservoir solution leads to condensation on the droplet. The humidity of the surrounding air has no effect on the observed phenomena as the droplet is isolated from the surrounding by the enclosed chamber. As the chamber is completely closed, all the vapor transport occurs between the droplet and the reservoir fluid.  There is no mass transfer out of the chamber. The experiment was conducted at room temperature of \(20 ^0 C\) and relative humidity of \(60 \, \%\). 

\begin{figure*}
\begin{center}
\includegraphics[width=1\textwidth]{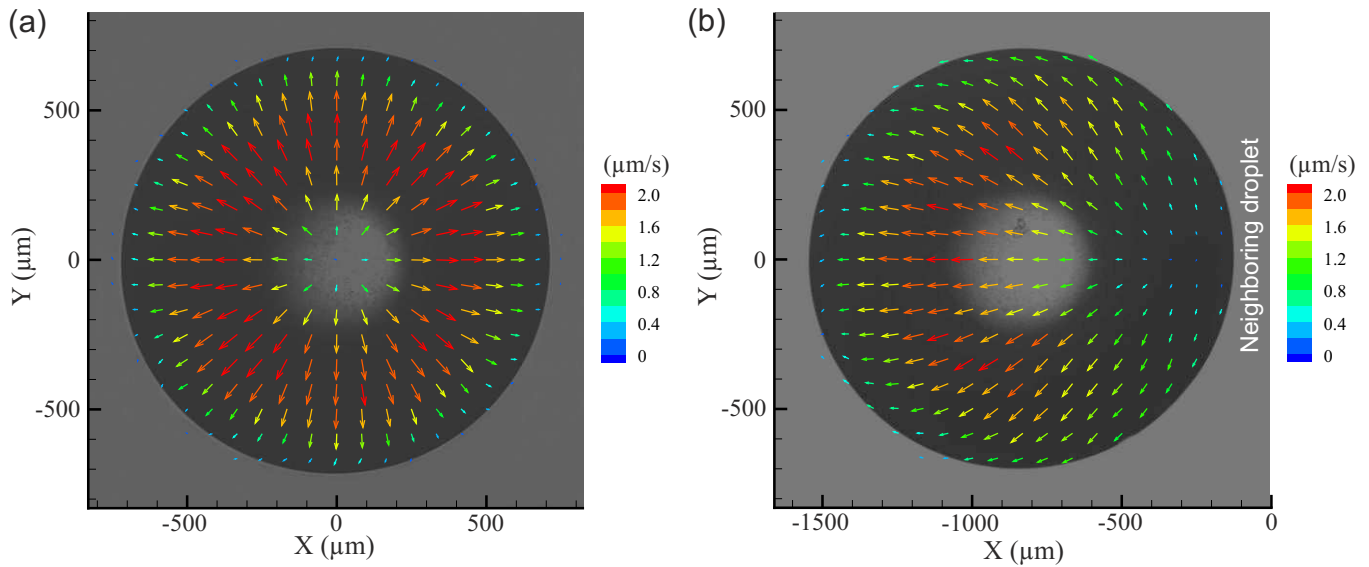}
\caption{\label{fig:combine} Velocity vector field at Z= 50 $\mu$m inside the condensing droplet (a) in the absence of a neighboring droplet and (b) with a neighboring droplet to the right.}
\end{center}
\end{figure*}

\subsection{Measurement of internal convection}
Confocal microscope (Leica) was used to capture images for PIV and Dynamic Studio V1.45 was used for processing of these images to find fluid velocity inside the droplet. Tracer particles were added to the solution of the droplet. In micro-PIV, the velocity of fluid is measured by tracing the tracer particles present in the fluid. The tracer particles used in this experiment are spherical fluorescent polystyrene particles of diameter ($d_p$) 2 $\mu$m (Invitrogen). The volume fraction of the tracer particles inside the droplet was taken equal to \(0.02 \, \% \). Laser source having wavelength of 488 nm was used to excite the fluorescent particles present inside the droplet. The fluorescent particles looks as bright spot inside the fluid by the excitation of the laser light. Fluorescent particles are used in the micro-PIV measurement for better signal to noise ratio. The tracer particles used in this experiment has a density ($\rho_p$) of \(1050 \, \textrm{kg/m}^3\). Difference in density between the tracer particles and the solution leads to gravity settling of the particles which can be estimated  from the settling velocity as \(U_g=\frac{d_p^2(\rho_p-\rho)}{18\mu}g\). The viscosity ($\mu$) of the solution is equal to \(1.09 \times 10^{-3} \; \textrm{Pa} \cdot \textrm{s}\) at 1 M concentration \cite{Kestin1981}. The density ($\rho$) of 1 M aqueous NaCl solution is equal to \(1038 \, \textrm{kg/m}^3\) \cite{Kenneth1984}. The calculated value of the settling velocity of the tracer particles for the present case is equal to \(2.3 \times 10^{-8} \, \textrm{m/s}\) which is negligible in comparison to the experimental value of the fluid velocity. The relaxation time of the tracer particles to the flow change is given by \(\tau_s=d_p^2\frac{\rho_p}{18\mu}\) which is equal to \(2.2 \times 10^{-7} \, \textrm{sec}\). For this small value of relaxation time, the particles quickly adjust to the velocity variation and follow the fluid flow without any lag.

\par
Use of a confocal microscope eliminates the volumetric illumination as it reconstruct the image by point scanning of the object. Use of pinhole in a confocal microscope eliminates background noise giving better contrast image. The image size was set equal to \(512 \times 512\) pixels. The time interval ($\Delta t$) between the images was kept equal to 2 sec. Velocity vector was obtained by processing the images using the PIV evaluation software (Dynamic Studio V1.45 ). Adoptive cross correlation was used with 2 steps refinement during the image processing. The interrogation area was set at \(32 \times 32\) pixels. There was \(25 \, \%\) interrogation area overlap to increase the vector resolution. Random movement of the particles due to Brownian effect may cause error in the velocity measurement. The error caused by the Brownian motion on the flow measurement was eliminated by averaging the velocity vector over five measurements \cite{Santiago1998,Meinhart1999}. The velocity measurements carried out in this experiment is 2D velocity measurement. Similar 2D velocity measurements were carried out at 14 X-Y planes having a vertical separation distance (\(\Delta z\)) of \(50 \, \mu m\) between the measurement planes. The flow observed in the present work shows almost steady pattern in a short period of time. The z component of velocity was calculated from these 2D velocity fields using steady continuity equation \cite{Pradhan2016,Pradhan2018a}. The velocity measurement was performed for both single and two droplets cases. The observed flow inside the two condensing droplets during interaction was compared with a single condensing droplet to find the effect of the neighboring droplet.  

\begin{figure*}
\begin{center}
\includegraphics[width=1\textwidth]{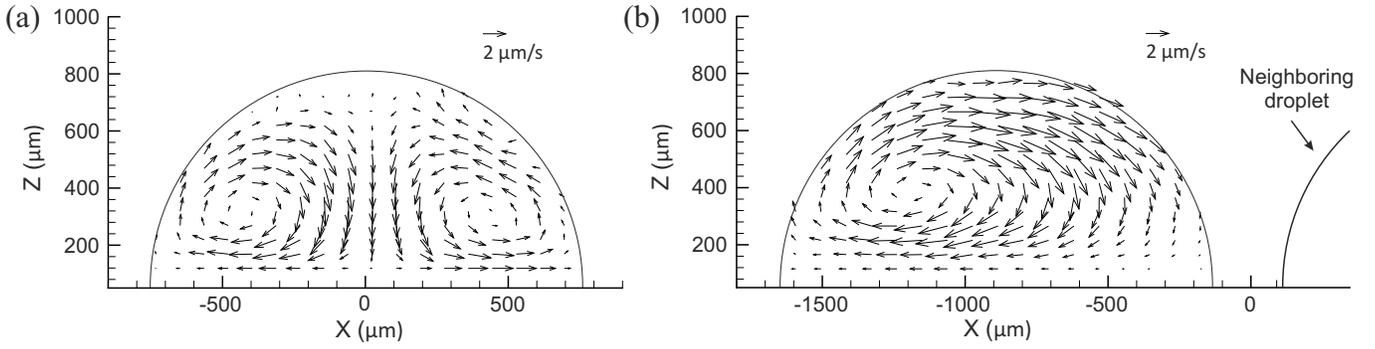}
\caption{\label{fig:3d} Reconstructed three dimensional velocity vector field at Y=0 inside the (a) single condensing droplet and (b) two interacting condensing droplets.}
\end{center}
\end{figure*}

\section{Results and discussion}
The velocity vector field inside a single condensing droplet is shown in Fig. \ref{fig:combine}(a) near the substrate surface at \(z=50 \mu m \) from the substrate. The velocity vector shows an outward flow towards the contact line and it shows a symmetric flow pattern. This flow pattern is same as that of flow pattern observed in our previous work on a condensing liquid droplet sandwiched between two parallel cover slips \cite{Pradhan2018b}. When another condensing droplet is placed to the right of the droplet, the fluid flow shows a change in pattern. Velocity vectors show no more symmetry pattern as observed in the single droplet where it shows outward flow from the proximity region of the two droplets (Fig. \ref{fig:combine}(b)).

\par
The velocity fields along a vertical plane at Y = 0 for single and two droplets configurations are shown in Fig. \ref{fig:3d}. The velocity vector field along the vertical plane is obtained by three dimensional velocity reconstruction from the velocity fields at 14 horizontal planes (XY plane) using continuity equation. Single droplet shows two symmetric circulating loops inside the droplet and in case of two droplets configuration, single circulating loop is observed in the vertical plane. 

\begin{figure}[htb!]
\begin{center}
\includegraphics[width=0.48\textwidth]{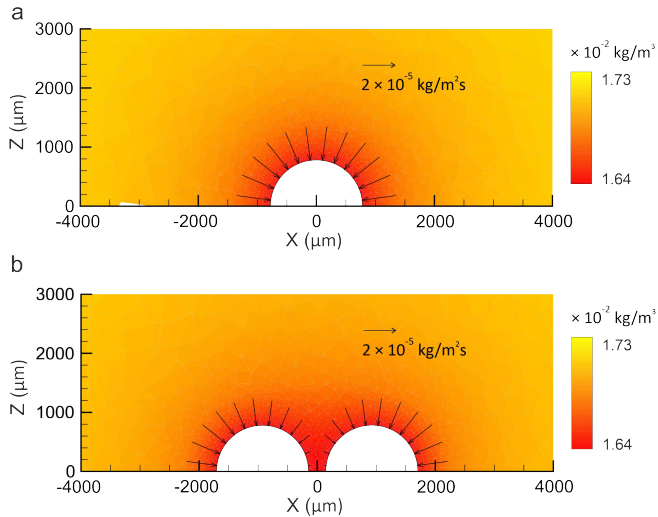}
\caption{\label{fig:flux} Vapor concentration distribution and condensation flux at the droplet surface for (a) single droplet and (b) two interacting droplets.}
\end{center}
\end{figure}

\par
The observed velocity vector field can be explained from the condensation flux calculation on the droplet surface. Vapor transport between the reservoir fluid (water) and the droplet occurs due to the vapor pressure difference at the liquid-air interface of the reservoir fluid and the droplet. Vapor concentration at the droplet surface (1 M) is given by Raoult's law as \(C_s = X_{solvent} C_{s0}\), where $X_{solvent}$ is the mole fraction of water in the solution. The saturated vapor concentration ($C_{s0}$) at the room temperature which is equal to  \(1.73 \times 10^{-2} \textrm{kg}/\textrm{m}^3\) obtained from CRC handbook \cite{crc}. The value of $C_{s}$ is equal to \(1.67 \times 10^{-2} \textrm{kg}/\textrm{m}^3\). The value of vapor concentration at the liquid-air interface of water (reservoir fluid) is same as $C_{s0}$. The value of $C_{s0}$ is greater than $C_s$. Hence, there is a vapor concentration gradient in the air medium enclosed inside the chamber. The transport of vapor ($C_v$) in the air medium is calculated from the diffusion equation:  

\begin{equation}\label{eq:dif}
\nabla ^2 C_v=0
\end{equation}

\par
The interface of the droplet has vapor concentration equal to $C_s$ and the reservoir fluid has a vapor concentration of $C_{s0}$ at its interface. The solid wall of the chamber is treated as no flux boundary condition. The diffusion equation is solved in the air domain surrounding the fluid using COMSOL. The condensation flux at the droplet interface is calculated as \(J=-D_v\Delta C_v \). The value of diffusion coefficient ($D_v$) of vapor is equal to \(2.4 \times 10^{-5} \, \textrm{m}^2/\textrm{s}\) \cite{crc}. The concentration of vapor around the droplet is shown in Fig. \ref{fig:flux}. The transport of water vapor in the air is dominated by quasi-steady diffusion process \cite{Hu2002,Sobac2011,Carle2013,Bhardwaj2018}. The condensation flux for both single condensing and two interacting droplets is shown in Fig. \ref{fig:flux}. Distribution of condensation flux around the droplet is uniform for single droplet.  The vapor concentration and condensation flux for two droplets case show asymmetric distribution pattern around the droplets. The accumulation of vapor at the proximity region suppresses the condensation flux at this region. The nature of the condensation flux around the droplet determines the nature of fluid flow inside the droplet.

\begin{figure*}
\begin{center}
\includegraphics[width=1\textwidth]{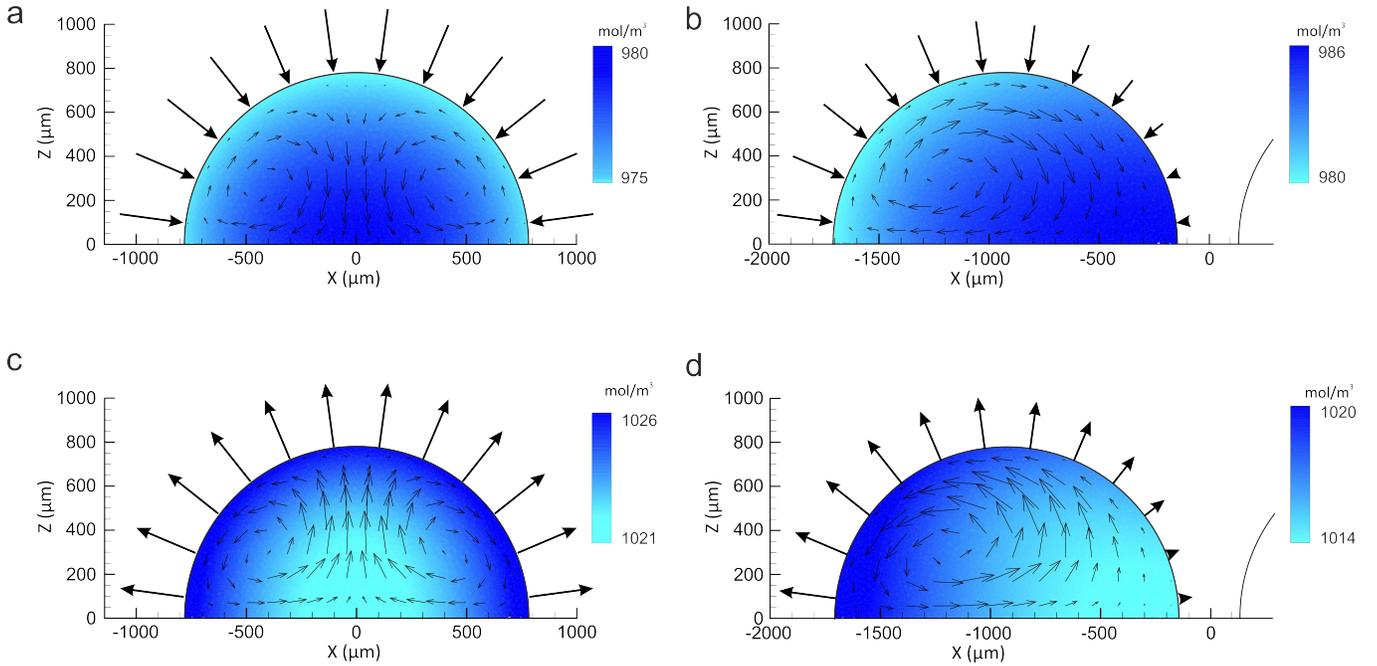}
\caption{\label{fig:mechanism} Flow mechanism inside the condensing droplets for (a) single condensing droplet, (b) two interacting condensing droplets, (c) single evaporating droplet and (d) two interacting evaporating droplets.}
\end{center}
\end{figure*}

\par
The capillary length is given by , \(\lambda_c = \sqrt{\sigma /\rho g } \). The value of surface tension ($\sigma$ ) for 1 M NaCl solution at $20 ^0$ C is equal to \(7.39 \times 10^{-2} \, \textrm{N/m}\) \cite{Ali2006}. The capillary length for the present case is equal to 2.7 mm. The droplet dimension in the present experiment (\(D= 1.56 \, mm\)) is below the capillary length. Hence, it is expected that the surface tension force is dominant over buoyancy force \cite{Cazabat2010,Xu2012,Barmi2014,Marin2016,Butzhammer2017,Malla2020}. However, recent experimental studies \cite{Pradhan2016,Kang2013,Lee2014,Edwards2018,Li2019} show that the internal flow pattern of aqueous droplet is dominated by buoyancy induced Rayleigh convection. The Marangoni convection is either absent or suppressed by Rayleigh convection. The cause of the absence or suppression of Marangoni convection is still not known. Therefore, the flow observed in the present case is described in terms of buoyancy effect and Marangoni effect is ignored. The internal convection of the droplets due to condensation is induced by buoyancy effect due to solute concentration variation in the droplet. The solute concentration ($c$) distribution in the droplet is obtained by numerically solving the continuity, momentum and species transport equations.

\begin{subequations}\label{eq:governing}
\begin{align}
        & \nabla \cdot \textbf{u} = 0 \\
        & \rho \frac{D \textbf{u}}{Dt} = - \nabla p + \mu \nabla^2 \textbf{u} + \rho g \\
        & \frac{D \textbf{c}}{Dt} = D_c \nabla^2 \textbf{c} 
\end{align}
\end{subequations}

\par
Here, $u$ and $p$ denote velocity and pressure respectively. Density ($\rho$) of the solution depends on solute concentration and \(\rho g\) in the momentum equation is the buoyancy body force responsible for inducing flow inside the droplet during condensation. The solid walls are taken as no slip and no flux boundary conditions. The liquid-air interface is taken as no-slip boundary \cite{Pradhan2016} as the tracer particles at the liquid-air interface do not show any movement. The boundary condition for species transport equation at the condensing interface is \(-D_c \nabla c \cdot \textbf{n} = -J \frac{c}{\rho_w}\). The value of solute diffusivity ($D_c$) for NaCl is equal to \(1.6 \times 10^{-9} \; m^2/s\) \cite{Riquelme2007} and water density ($\rho_w$) is equal to \(998 \, \textrm{kg/m}^3\). The equations are solved using COMSOL. The details numerical procedure are same as that of evaporating droplets as explained by \citet{Pradhan2016}. 

\begin{figure}[htb!]
\begin{center}
\includegraphics[width=0.48\textwidth]{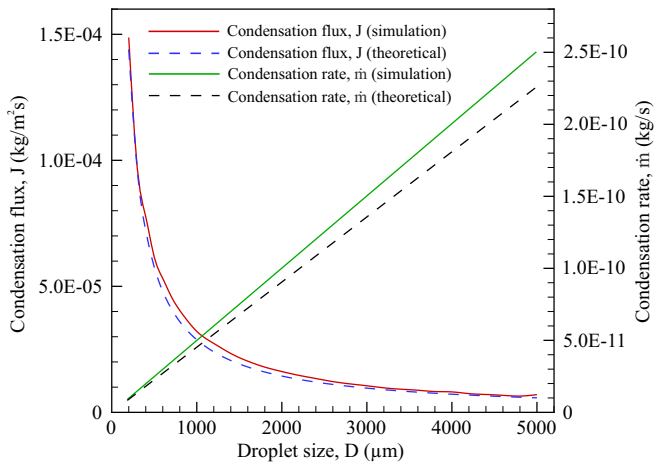}
\caption{\label{fig:sizecond} Effect of droplet diameter on the condensation flux for single droplet.}
\end{center}
\end{figure}

\par
The solute concentration distribution and velocity field obtained from the simulation are presented in Fig. \ref{fig:mechanism} to describe the flow physics. Condensation at the droplet surface lowers the solute concentration near the interface region as shown in Figs. \ref{fig:mechanism}(a) and (b). Fluid with low density (low concentration) moves upward along the interface leading to circulating flow inside the droplet which is symmetric in nature due to the symmetric condensation flux for single droplet. There is a suppression of condensation flux at the proximity region of the two droplets.  Less condensation at this region leads to higher solute concentration at the proximity region of the two droplets as compared to other regions. The solute concentration at the opposite region is very less due to higher condensation flux. Hence, fluid moves upward direction at this point due to buoyancy and the fluid near the proximity region shows downward movement creating a circulating loop.

\par
The observed flow pattern for condensing droplet has been compared with the evaporating droplet \cite{Pradhan2016} keeping the magnitude of evaporative flux same as that of the magnitude of condensation flux in the simulation. Figs. \ref{fig:mechanism}(c) and (d) show the flow pattern for single and interacting droplets during evaporation. The flow pattern observed for single condensing droplet (Fig. \ref{fig:mechanism}(a)) has an opposite behavior in comparison to the flow pattern in case of evaporating droplet (Fig. \ref{fig:mechanism}(c)). Similarly, the direction of single circulation loop for two interacting condensing droplets  (Fig. \ref{fig:mechanism}(b)) is opposite to the two interacting evaporating droplets (Fig. \ref{fig:mechanism}(d)). Also it is observed from the simulation that the strength of velocity for condensing droplet is less as compared to evaporating droplet. Maximum velocity for condensing droplet (Fig. \ref{fig:mechanism}(a)) is equal to 3.5 $\mu$m where as it is 5.0 $\mu$m for evaporating droplet (Fig. \ref{fig:mechanism}(c)). This can be explained from the solute concentration distribution. In case of condensing droplet, low concentration fluid due to condensation occupies at the top of the droplet causing stratification. On the other hand, higher concentration fluid develops at the interface due to evaporation in the evaporating droplet and the higher density fluid at the top slides down due to buoyancy enhancing the flow.  

\begin{figure}[htb!]
\begin{center}
\includegraphics[width=0.47\textwidth]{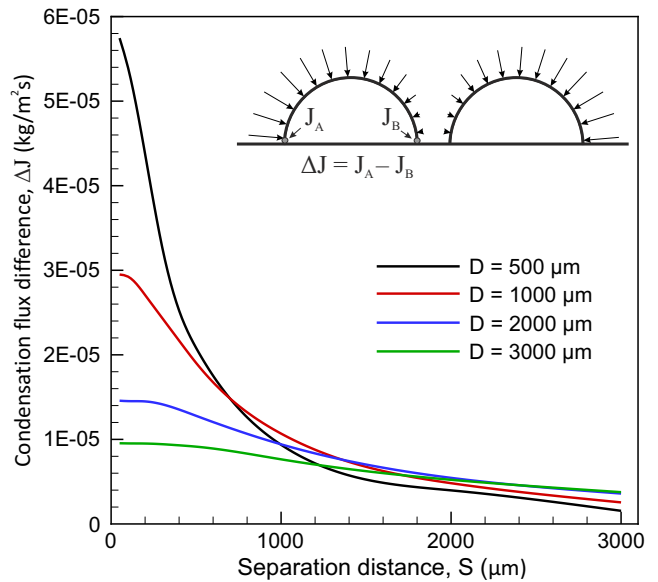}
\caption{\label{fig:size}  Difference in condensation flux ($\Delta J$) on both the sides of the droplet with separation distance.}
\end{center}
\end{figure}

\par
Condensation flux ($J$) for single droplet configuration is reported in Fig. \ref{fig:sizecond} at different droplet size. It shows an asymptotically decrease in condensation flux with increase in droplet diameter. The evaporative flux for an evaporating hemispherical droplet is inversely proportional to the droplet diameter (\(J \propto 1/D\)) as studied by \citet{Hu2002}. Similar behavior can be expected for condensing droplet if the condensation process is quasi-steady and diffusion driven which is the case in the present study. Therefore, condensation flux on the droplet surface can be given by \cite{Hu2002} 

\begin{equation}\label{eq:flux}
J=\frac{2D_v (C_{s0}-C_{s})}{D}
\end{equation}

\par
The equation is usually used for calculating evaporative flux from a hemispherical droplet (contact angle = 90 $^0$) evaporating in an ambient environment. In this equation, the ambient vapor concentration is replaced by the reservoir vapor concentration ($C_{s0}$). The condensation flux obtained from the equation (Eq. \ref{eq:flux}) matches with the the condensation flux obtained from simulation. The total condensation flux on the droplet surface can be obtained by multiplying the condensation flux (Eq. \ref{eq:flux}) with the exposed surface area ($A$) and given by

\begin{equation}\label{eq:totalflux}
\dot{m}= AJ= \pi D D_v (C_{s0}-C_{s})
\end{equation}

\par
The equation shows total condensation rate is directly proportional to the diameter of the droplet. Total condensation rate obtained from simulation is shown in Fig. \ref{fig:sizecond} which has a linear variation with droplet size. Condensation rate obtained from the equation \ref{eq:totalflux} and simulation deviates with increase in size. Small deviation in condensation flux gets magnified with droplet size when the condensation flux is multiplied with the exposed surface area. Though the reservoir fluid (water) only occupies the annular grove of the chamber, the inside environment of the chamber behaves as if the droplets are surrounded by an ambient vapor cloud of concentration equal to the saturated vapor concentration of reservoir interface ($C_{s0}$).

\begin{figure*}
\begin{center}
\includegraphics[width=1\textwidth]{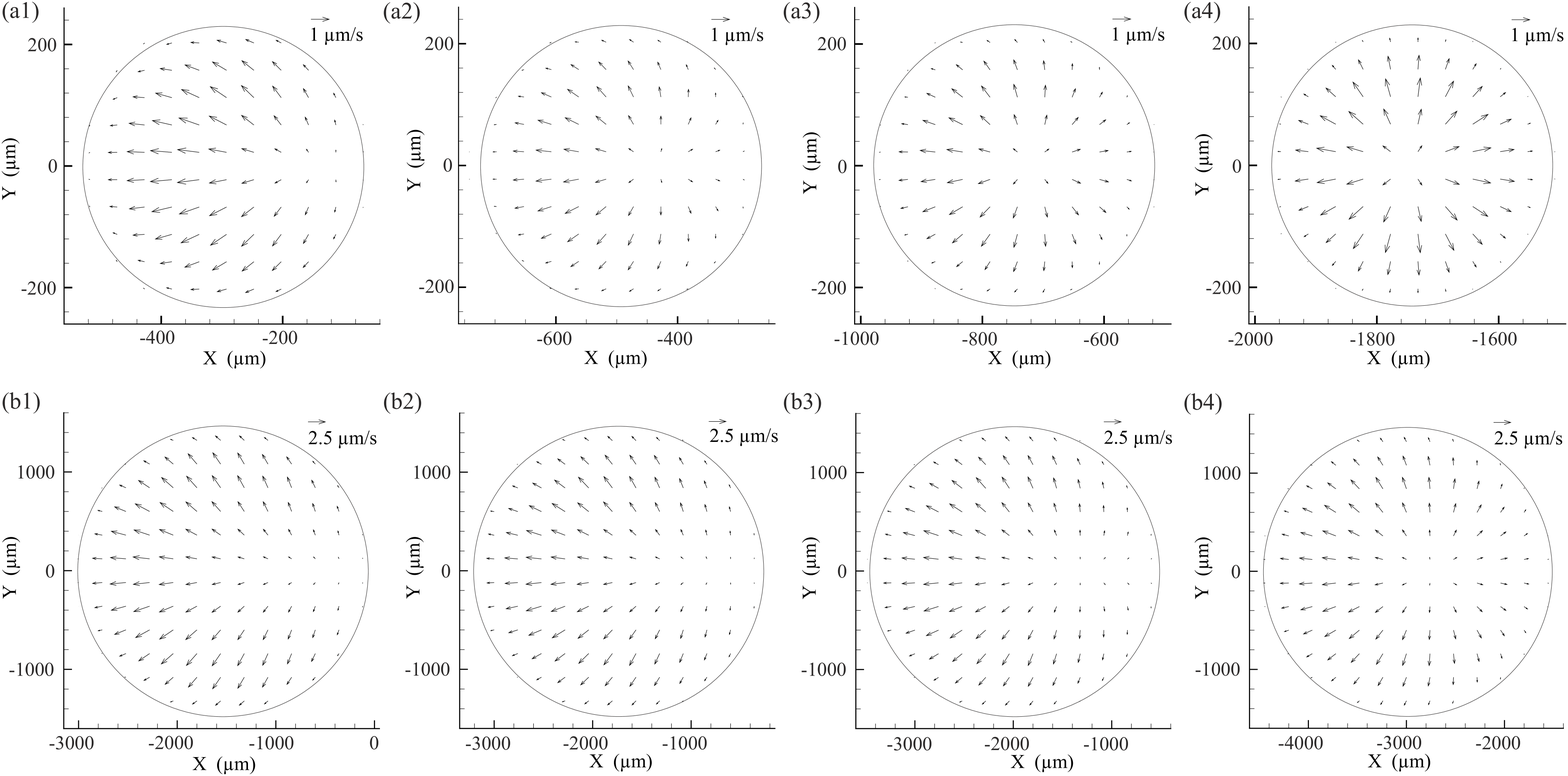}
\caption{\label{fig:distance} Velocity vector field in XY plane at Z= 50 $\mu$m inside the condensing droplet having droplet size of 500 $\mu$m (a1-a4) and 3000 $\mu$m (b1-b4). The separation distance between the droplets is 100 $\mu$m (a1 $\&$ b1), 500 $\mu$m (a2 $\&$ b2), 1000 $\mu$m (a3 $\&$ b3) and 3000 $\mu$m (a4 $\&$ b4).}
\end{center}
\end{figure*}

\par 
The effect of the neighboring droplet is felt due to the modification of condensation flux around the droplet. The condensation flux distribution around the droplet determines the solute concentration distribution which governs the fluid flow. The effect of the separation distance on the flow pattern can indirectly be studied by investigating the effect on the condensation flux distribution. The difference of condensation flux (\(\Delta J \)) at the two ends of the droplet in the presence of the neighboring droplet is presented in Fig. \ref{fig:size}. It shows an asymptotically decrease in the difference of condensation flux with separation distance. This indicates the asymmetric nature of the flow pattern decreases with the increase in separation distance. Also it is observed that the difference in condensation flux is lower for larger droplet size. This can be explained from  Fig. \ref{fig:sizecond} that higher droplet size gives less condensation flux which also leads to smaller value in difference in condensation flux. Hence, it is expected that more asymmetric in flow pattern is observed for smaller droplet size as compared to larger droplet size. However, the fluid flow inside the droplet depends on droplet size and cumulative condensation rate along with condensation flux. Hence, asymmetric nature of flow with respect to different size can not be interpreted from the asymmetric condensation flux with droplet size.

\begin{figure}[htb!]
\begin{center}
\includegraphics[width=0.49\textwidth]{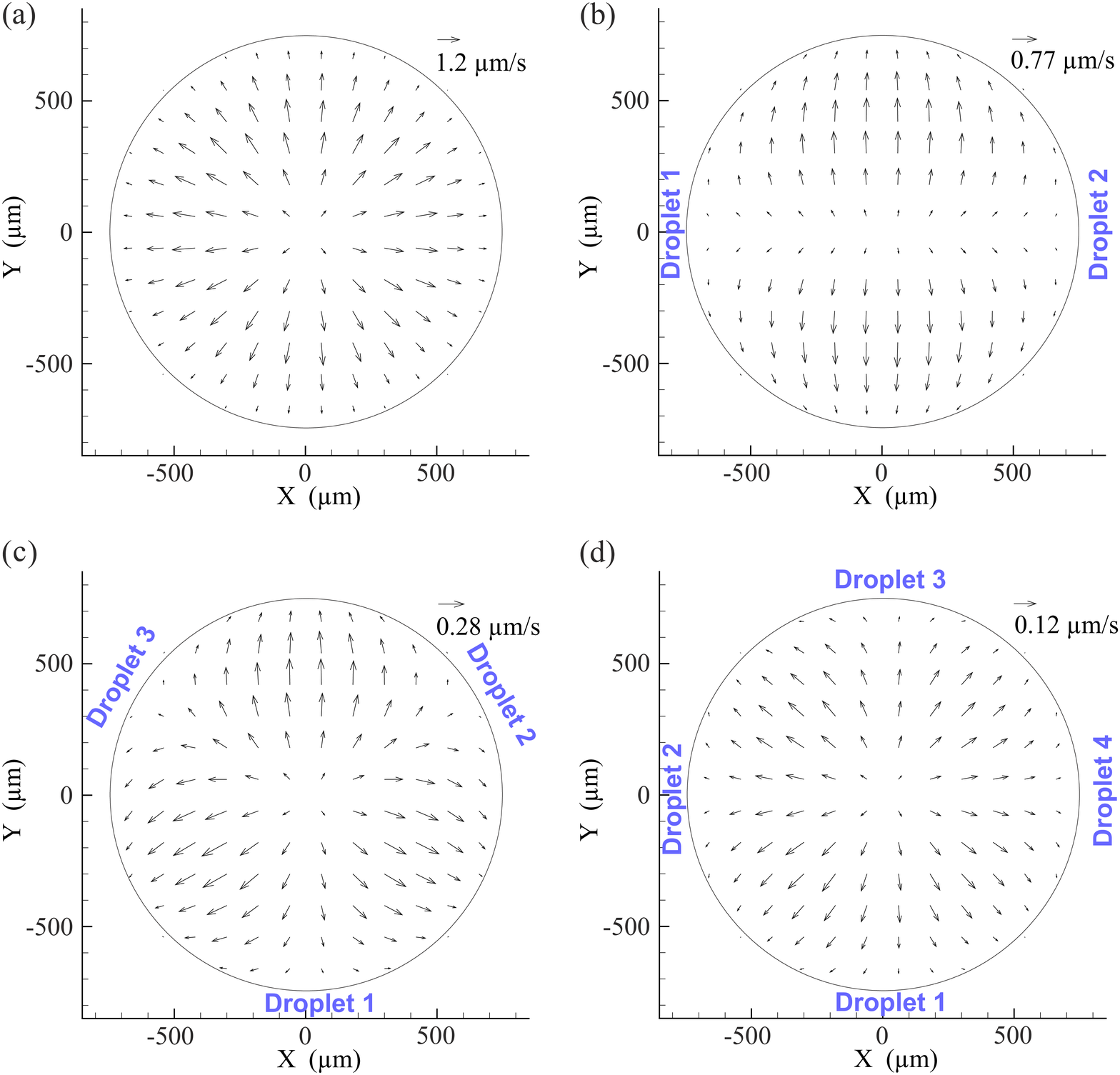}
\caption{\label{fig:multiple}  Velocity vector fields in XY plane at Z= 50 $\mu$m inside the condensing droplet when symmetrically surrounded by (a) no droplets, (b) two droplets on both the sides, (c) three droplets at 120 $^0$ from each other and (d) four droplets at 90 $^0$ from each other. The neighboring droplets are separated by a separation distance of 290 $\mu$m from the central droplet.}
\end{center}
\end{figure}

\par
The velocity vector fields at different separation distance is presented in Fig. \ref{fig:distance} for droplet diameters of 500 $\mu$m and 3000 $\mu$m. It is difficult to place two neighboring droplets at a fixed particular separation distance. Hence, we studied the effect of separation distance by simulation which provides better flexibility to study the behavior at a wide range of separation distance.  It shows the asymmetric nature of flow pattern deceases with increase in separation distance. This is because of the decrease in evaporative flux difference with separation distance. Similar trend is observed in both the size of droplets. However, the influence of neighboring droplet is felt in longer separation distance for larger droplet as compared to smaller droplet. This can be seen from Figs. \ref{fig:distance}(a4) and (b4) where for same separation distance of 3000 $\mu$m, the flow pattern for smaller drop (D=500 $\mu$m) is almost symmetric. However, at the same separation distance, flow pattern for larger droplet (D=3000 $\mu$m) has still asymmetric behavior.

\par
The flow pattern inside the droplet also depends on the number of neighboring droplets. The velocity field in X-Y plane at Z= 50 $\mu$m is presented in Fig. \ref{fig:multiple} when the droplet is surrounded by neighboring droplets at different configurations. The results are obtained from simulation. The diameter of the droplets is taken equal to 1.56 mm which is same as that of experimental value. The droplet is symmetrically surrounded by neighboring droplets. The velocity vector fields show that when the number of neighboring droplets increases, the symmetric nature of fluid flow reduces and the flow pattern approaches towards single droplet configuration. The flow pattern inside droplet symmetrically surrounded by four droplets (Fig. \ref{fig:multiple}(d)) is almost same as that of single droplet without a neighboring droplet (Fig. \ref{fig:multiple}(a)). However, the flow strength in  Fig. \ref{fig:multiple}(d) is one tenth of the flow strength of the single droplet case in Fig. \ref{fig:multiple}(a). This is because of the suppression of the condensation rate in the presence of neighboring droplets.

\section{Conclusion}
Internal convection of vapor mediated sensing of two condensing droplets has been studied experimentally using confocal micro-PIV. The two condensing droplets sense the presence of other through vapor mediated interaction without having any physical contact between the two droplets. Condensation on the aqueous droplets is carried out inside a closed chamber where the droplets are placed on a circular hydrophobic glass cover slip which is surrounded by a reservoir of water. Condensation on the droplets occurs owing to the difference in vapor pressure between the droplet and the surrounding reservoir fluid (water). Condensation at the droplet interface generates concentration gradient inside the droplet inducing density driven buoyancy convection. The flow pattern for a single condensing droplet shows symmetric circulation pattern which is opposite to the direction of flow in case of evaporating droplet. A neighboring droplet influences the flow structure inside the condensing droplet. The flow alternation is because of the change in condensation flux distribution at the droplet interface due to its vapor mediated interaction with the neighboring condensing droplet. The internal convection observed in case of two condensing droplets shows opposite flow direction as observed in case of two evaporating droplets. The asymmetric nature of the flow pattern reduces as the  separation distance between the two droplets increases. The fluid flow pattern inside the droplet is also depends on the number of neighboring droplets. The asymmetric nature of the flow reduces with increase in the number of neighboring droplets.

\begin{acknowledgments}
Authors thank the DST, Government of India and IIT Kanpur for the financial support to conduct the research. The work has been carried out at Microfluidics and Sensor Laboratory, IIT Kanpur.
\end{acknowledgments}

\nocite{*}
\bibliography{reference}

\end{document}